\documentstyle[aps,prd]{revtex}

\newcommand{\be}{\begin{equation}}
\newcommand{\ee}{\end{equation}}
\newcommand{\bea}{\begin{eqnarray}}
\newcommand{\eea}{\end{eqnarray}}
\newcommand{\pa}{\partial}
\newcommand{\ve}{\varepsilon}

\begin{document}

\title{Lapse function for maximally sliced Brill-Lindquist initial data}

\author{Piotr Jaranowski
\\{\it Institute of Theoretical Physics, University of Bia{\l}ystok}\\
  {\it Lipowa 41, 15-424 Bia{\l}ystok, Poland}
\thanks{Email address: pio@alpha.uwb.edu.pl}\vspace{2ex}\\
Gerhard Sch\"afer
\\{\it Theoretisch-Physikalisches Institut, 
       Friedrich-Schiller-Universit\"at}\\
  {\it Max-Wien-Platz 1, 07743 Jena, Germany}
\thanks{Email address: gos@tpi.uni-jena.de}}


\maketitle

\begin{abstract}

For binary black holes the lapse function corresponding to the Brill-Lindquist
initial value solution for uncharged black holes is given in analytic form under
the maximal slicing condition.  In the limiting case of very small ratio of mass
to separation between the black holes the surface defined by the zero value of
the lapse function coincides with the minimal surfaces around the singularities.

\vspace{0.3cm}\noindent PACS number(s):  04.70.Bw, 04.20.Ex, 04.20.Fy
\end{abstract}

\vspace{0.5cm}

In a recent paper \cite{JS00}, the authors re-obtained the Brill-Lindquist
initial value solution for two electrically uncharged black holes \cite{BL63}
using a point-mass model for the black holes followed by Hadamard's ``partie
finie'' regularization procedure in 3-dimensional space.  (It can be shown
that the same result appears by applying the dimensional regularization of
Ref.\ \cite{DJS}.)  In the present paper we supplement this result by the
explicit calculation of the initial lapse function.  By the time-symmetry
assumption in the Brill-Lindquist initial value solution, the vanishing of the
initial shift function automatically occurs.  Therefore, all initial value
metric coefficients will be known.  Our result, for the first time, allows the
determination of the motion of test fields and particles near the initial
hypersurface for time scales which are small against the time scale of the
motion of the two black holes.  Similarly, the result will allow discussions
of quantum gravitational problems near the initial hypersurface.  Furthermore,
the knowledge of the initial 4-metric is a step towards an (approximate)
analytic treatment of the motion of two black holes starting from
Brill-Lindquist initial value data.  Finally, numerical head-on-collision
simulations starting from those initial data may compare the numerically
obtained initial lapse function with the analytic one of the present paper to
check the accuracy of the numerical code.  In our brief report, we
additionally will address the question of how the surface of vanishing initial
lapse function, which is the surface of infinite redshift for observers
momentarily at rest in the initial hypersurface, is related in some limiting
case to the minimal 2-surfaces associated with the Einstein-Rosen bridges of
the Brill-Lindquist 3-geometry (see Ref.\ \cite{BL63}), knowing that for the
Schwarzschild black hole geometry the both types of surfaces coincide (see,
e.g., Ref.\ \cite{MTW}).

We use units in which $16\pi G=c=1$, where $G$ is the Newtonian gravitational
constant and $c$ is the velocity of light. In the paper all vectors and their
lengths are defined in the 3-dimensional Euclidean space endowed with a
standard Euclidean metric; ${\bf x}$ is the position of an arbitrary point in
this space.
 
In the generalized isotropic ADM gauge [see Eqs.\ (7-4.22) and (7-4.23) in Ref.\ 
\cite{ADM}] the 3-metric takes the form
\be
\label{eq1}
g_{ij} \equiv \left(1+\frac{1}{8}\phi\right)^4 \delta_{ij} + h^{\text{TT}}_{ij}
\ee
and for the canonical conjugate $\pi^{ij}$ of the 3-metric, which in terms of 
the extrinsic curvature $K_{ij}$ of the spacelike hypersurfaces
$t={\text{const}}$ is given by (see, e.g., Ref.\ \cite{MTW})
\be
\label{eq2}
\pi^{ij} = - g^{1/2} (g^{il}g^{jm} - g^{ij}g^{lm})  K_{lm},    
\ee
where $g^{ij}$ is the inverse of $g_{ij}$ and $g\equiv{\det(g_{ij})}$,
the condition
\be
\label{eq3}
\pi^{ii} \equiv 0  
\ee
holds. 

In case of time-symmetric initial conditions the black-hole linear momenta 
$p_{ai}$ ($a=1,2$) and field canonical conjugate $\pi^{ij}$, and thus the 
shift function, $N_i=g_{0i}$, too, are zero
\be
\label{eq4}
p_{ai} = 0, \quad \pi^{ij} = 0, \quad N_i =0.
\ee
Conformal flatness, i.e.
\be
\label{eq5}
g_{ij} = \left(1+\frac{1}{8}\phi\right)^4 \delta_{ij},
\ee
additionally implies the vanishing of the transverse-traceless part of the 
3-metric,
\be
\label{eq6}
h^{\text{TT}}_{ij}=0.
\ee
In this case, the coordinate condition (\ref{eq3}) coincides with the maximal
slicing condition, i.e., $g^{ij}K_{ij}=0$.
Under the above conditions, the lapse function $N_0$ 
($N_0^2=-g_{00}+g^{ij}N_iN_j=-g_{00}$, as $N_i=0$) fulfills the equation
\be
\label{eq7}
g^{1/2} g^{ij} D_iD_jN_0 = \frac{1}{4} N_0\, g^{1/2}R,
\ee
where $D_i$ means the covariant derivative with respect to $g_{ij}$, and $R$ is 
the intrinsic curvature of the hypersurface $t={\text{const}}$. The Eq. 
(\ref{eq7}) is easily obtained from the field equation for $\pi^{ii}$, i.e., 
from ${\partial \pi^{ii}}/{\partial t}\equiv0$ [see, e.g., Eq.\ (2.7a) in Ref.\ 
\cite{S85}].

After substituting the ansatz 
\be
\label{eq8}
N_0 = \frac{1-\psi/8}{1+\phi/8}
\ee
into the Eq.\ (\ref{eq7}), we obtain the following equation for the function 
$\psi$: 
\be
\label{eq9}
\left(1+\frac{1}{8}\phi\right) \Delta\psi
= \left(1-\frac{1}{8}\psi\right) \Delta\phi.
\ee
To derive Eq.\ (\ref{eq9}) we have employed the relation, valid for
application on scalar functions,
$$
g^{1/2} g^{ij} D_i D_j = \frac{\pa}{\pa x^i}
\left( g^{1/2} g^{ij} \frac{\pa}{\pa x^j} \right)
$$
and the following formula, valid when $h^{\text{TT}}_{ij}=0$ (see, e.g., Ref.\ 
\cite{S85}):
$$
g^{1/2} R = -\left(1+\frac{1}{8}\phi\right) \Delta\phi.
$$

The Brill-Lindquist solution reads, \cite{BL63},
\be
\label{eq10}
\phi = 8 \left(\frac{\alpha_1}{r_1}+\frac{\alpha_2}{r_2}\right),
\ee
where the parameters $\alpha_1$ and $\alpha_2$ are constants and where $r_1$ and 
$r_2$ denote the coordinate distances between the field point and the black hole 
positions in the conformally related flat space. The parameters $\alpha_1$,
$\alpha_2$ are related to the bare masses $m_1$, $m_2$ of the two black holes
through the formulas (see Appendix~A in Ref.\ \cite{JS99})
\begin{mathletters}
\label{alpha-m}
\bea
\alpha_1 &=& -\frac{1}{4}\left(2r_{12}+\frac{m_2-m_1}{16\pi}\right)
+ \frac{1}{4}r_{12} \sqrt{4+\frac{m_1+m_2}{4\pi r_{12}}
+\left(\frac{m_1-m_2}{16\pi r_{12}}\right)^2},
\\[2ex]
\alpha_2 &=& -\frac{1}{4}\left(2r_{12}+\frac{m_1-m_2}{16\pi}\right)
+ \frac{1}{4}r_{12} \sqrt{4+\frac{m_1+m_2}{4\pi r_{12}}
+\left(\frac{m_1-m_2}{16\pi r_{12}}\right)^2},
\eea
\end{mathletters}
where $r_{12}$ denotes the coordinate distance of the two black holes.

Equation (\ref{eq9}) is invariant against $\psi\rightarrow-\phi$, 
$\phi\rightarrow-\psi$, which obviously suggests identical structures of the 
functions $\phi$ and $\psi$ involved in this equation. We therefore search
for a solution of the function $\psi$ of the form
\be
\label{eq11}
\psi = 8 \left(\frac{\beta_1}{r_1}+\frac{\beta_2}{r_2}\right),
\ee
where $\beta_1$ and $\beta_2$ are some new constants.
After substitution Eqs.\ (\ref{eq10}) and (\ref{eq11}) into Eq.\ (\ref{eq9}) we
make use of the regularization procedure of Ref.\ \cite{JS00} (or Ref.\ 
\cite{DJS}) which relies here on replacing $f({\bf x})\delta({\bf x}-{\bf x}_a)$ 
by $f_{\rm reg}({\bf x}_a)\delta({\bf x}-{\bf x}_a)$, where the regularized 
value $f_{\rm reg}({\bf x}_a)$ of the function $f$ at its (possibly) singular 
point ${\bf x}={\bf x}_a$ we define by means of the Hadamard's ``partie finie'' 
procedure. We expand $f\left({\bf x}_a+\ve{\bf n}\right)$ (where ${\bf n}$ is a 
unit vector and $\ve>0$ is a number) into a Laurent series around $\ve=0$ and as 
the regularized value of the function $f$ at ${\bf x}_a$ we take the zero-order 
coefficient of the series averaged over all directions ${\bf n}$:
\be
\label{hpf1}
f\left({\bf x}_a+\ve{\bf n}\right)
= \sum\limits_{m=-N}^{\infty} a_m({\bf n})\,\ve^m, \quad
f_{\rm reg}\left({\bf x}_a\right)
\equiv \frac{1}{4\pi}\oint\!d\Omega\,a_0({\bf n}).
\ee
The usage of the definition (\ref{hpf1}) boils down in our computation to simple 
formulas
\begin{mathletters}
\label{hpf2}
\bea
\frac{\delta({\bf x}-{\bf x}_1)}{|{\bf x}-{\bf x}_1|} = 0,\quad
\frac{\delta({\bf x}-{\bf x}_1)}{|{\bf x}-{\bf x}_2|}
= \frac{\delta({\bf x}-{\bf x}_1)}{|{\bf x}_1-{\bf x}_2|},
\\[2ex]
\frac{\delta({\bf x}-{\bf x}_2)}{|{\bf x}-{\bf x}_2|} = 0,\quad
\frac{\delta({\bf x}-{\bf x}_2)}{|{\bf x}-{\bf x}_1|}
= \frac{\delta({\bf x}-{\bf x}_2)}{|{\bf x}_1-{\bf x}_2|}.
\eea
\end{mathletters}

Making use of Eqs.\ (\ref{hpf2}) we have obtained, by comparing the coefficients 
at Dirac delta functions, the following algebraic equations for the constants 
$\beta_1$ and $\beta_2$:
\begin{mathletters}
\bea
\left(1 + \frac{\alpha_2}{r_{12}}\right) \beta_1
&=& \left(1 - \frac{\beta_2}{r_{12}}\right) \alpha_1, 
\\[2ex]
\left(1 + \frac{\alpha_1}{r_{12}}\right) \beta_2
&=& \left(1 - \frac{\beta_1}{r_{12}}\right) \alpha_2. 
\eea
\end{mathletters}
The unique solution of the both equations is easily obtained. It reads
\begin{mathletters}
\label{beta}
\bea
\beta_1 &=& \alpha_1 \frac{r_{12}+\alpha_1-\alpha_2}{r_{12}+\alpha_1+\alpha_2},
\\[2ex]
\beta_2 &=& \alpha_2 \frac{r_{12}+\alpha_2-\alpha_1}{r_{12}+\alpha_1+\alpha_2}.
\eea
\end{mathletters}
In the limit $r_{12}\to\infty$, where two completely separated
Schwarzschild black holes have to arise, it holds: $\beta_1=\alpha_1$ and 
$\beta_2=\alpha_2$. If $\alpha_1=0$, then $\beta_1=0$ and $\beta_2=\alpha_2$ 
(analogously, if $\alpha_2=0$, then $\beta_2=0$ and $\beta_1=\alpha_1$). Also, 
if $\alpha_1=\alpha_2$, then $\beta_1=\beta_2$. Thus, the solution for $\psi$ 
does have correct limiting values. Furthermore, the post-Newtonian expansion of 
our solution for $\psi$ coincides with the corresponding part of the
lapse function obtained in Ref.\ \cite{S85} through second post-Newtonian order.

For completeness, we give the initial 4-dimensional line element. It takes the 
form 
\be
\label{eq15}
ds^2 = - \left(\frac{1- \psi/8}{1+ \phi/8}\right)^2 dt^2
+ ({1+ \phi/8})^4 (dx^2 + dy^2 + dz^2),
\ee
where $x$, $y$, $z$ denote Cartesian coordinates and where the Eqs.\ 
(\ref{eq10}), (\ref{eq11}), and (\ref{beta}) [and also Eqs.\ (\ref{alpha-m}) if 
the dependence in terms of bare masses is needed] have to be substituted.

In the case of Schwarzschild black holes (there, $\psi=\phi$), the surface
$N_0=0$ (being there the cross-section of a $t={\text{const}}$
slice with the  event
horizon) coincides with the throat of the Einstein-Rosen bridge which is a
minimal 2-surface in 3-space.  For binary black-hole systems, the surfaces
$N_0=0$ need not to be identical with minimal 2-surfaces, respectively throats,
in 3-space.  However, an explicit calculation has
shown that the initial value surface $N_0=0$ (or, $\psi = 8$), coincides with
the minimal surfaces around the singularities in the limiting case of very large
separation between the black holes, i.e., we exactly have reproduced the Eq.\
(27) in Ref.\ \cite{BL63} for uncharged black holes.

\bigskip

P.J.\ thanks the Theoretisch-Physikalisches Institut of the FSU Jena for
hospitality during the realization of this work.  The work of P.J.\ was
supported by the EU Programme ``Improving the Human Research Potential and the
Socio-Economic Knowledge Base'' (Research Training Network Contract
HPRN-CT-2000-00137) and by the Polish KBN Grant No.\ 5 P03B 034 20.

\end{document}